\documentstyle[prl,aps,floats,psfig]{revtex}

\draft
\preprint{NEVIS-1529}
\begin{document}

\vspace{-1.2cm}
\title{\begin{flushright} {\rm NEVIS-1529} \\ \end{flushright} A High Statistics Search for $\nu_\mu(\overline\nu_\mu) \rightarrow
\nu_e(\overline\nu_e)$ Oscillations in the Small Mixing Angle Regime}
\author{
        A.~Romosan,$^2$ C.~G.~Arroyo,$^2$ L.~de~Barbaro,$^5$
        P.~de~Barbaro,$^7$ A.~O.~Bazarko,$^2$ R.~H.~Bernstein,$^3$
        A.~Bodek,$^7$ T.~Bolton,$^4$ H.~Budd,$^7$ J.~Conrad,$^2$
        R.~B.~Drucker,$^6$ D.~A.~Harris,$^7$ R.~A.~Johnson,$^1$
        J.~H.~Kim,$^2$ B.~J.~King,$^2$ T.~Kinnel,$^8$ M.~J.~Lamm,$^3$
        W.~C.~Lefmann,$^2$ W.~Marsh,$^3$ K.~S.~McFarland,$^3$
        C.~McNulty,$^2$ S.~R.~Mishra,$^2$ D.~Naples,$^4$
        P.~Z.~Quintas,$^2$ W.~K.~Sakumoto,$^7$ H. Schellman,$^5$
        F.~J.~Sciulli,$^2$ W.~G.~Seligman,$^2$ M.~H.~Shaevitz,$^2$
        W.~H.~Smith,$^8$ P.~Spentzouris,$^2$ E.~G.~Stern,$^2 $
        M.~Vakili,$^1$ U.~K.~Yang,$^7$ and J.~Yu$^3$
}
\address{
$^1$ University of Cincinnati, Cincinnati, OH 45221 \\
$^2$ Columbia University, New York, NY 10027 \\
$^3$ Fermi National Accelerator Laboratory, Batavia, IL 60510 \\
$^4$ Kansas State University, Manhattan, KS 66506 \\
$^5$ Northwestern University, Evanston, IL 60208 \\
$^6$ University of Oregon, Eugene, OR 97403 \\
$^7$ University of Rochester, Rochester, NY 14627 \\
$^8$ University of Wisconsin, Madison, WI 53706 \\
}
\date{\today}
\maketitle
\begin{abstract}
Limits on $\nu_\mu (\overline{\nu}_\mu) \to \nu_e (\overline{\nu}_e)$
oscillations based on a statistical separation of $\nu_e N$ charged
current interactions in the CCFR detector at Fermilab are presented.
$\nu_e$ interactions are identified by the difference in the
longitudinal shower energy deposition pattern of $\nu_e N \rightarrow
eX$ versus $\nu_\mu N \rightarrow \nu_\mu X$ interactions. Neutrino
energies range from 30 to 600 GeV with a mean of 140 GeV, and
$\nu_\mu$ flight lengths vary from 0.9 km to 1.4 km. The lowest 90\%
confidence upper limit in $\sin^2 2\alpha$ of $1.1 \times 10^{-3}$ is
obtained at $\Delta m^2 \sim 300$~${\rm eV^2}$. For $\sin^2 2\alpha =
1$, $\Delta m^2 > 1.6$~${\rm eV^2}$ is excluded, and for $\Delta m^2
\gg 1000$~${\rm eV^2}$, $\sin^2 2\alpha > 1.8 \times 10^{-3}$ is
excluded. This result is the most stringent limit to date for $\Delta
m^2 > 25$~${\rm eV^2}$ and it excludes the high $\Delta m^2$
oscillation region favoured by the LSND experiment. The
$\nu_\mu$-to-$\nu_e$ cross-section ratio was measured as a test of
$\nu_\mu (\bar\nu_\mu) \leftrightarrow \nu_e (\bar\nu_e)$ universality
to be $1.026 \pm 0.055$.
\end{abstract}
\twocolumn

The existence of neutrino mass and mixing would have important
implications for fundamental problems in both particle physics and
cosmology. These include violation of lepton family number
conservation, the mass of the universe, and the observed neutrino
deficits from the sun and from atmospheric sources. Neutrino
oscillations are a necessary consequence of non-zero neutrino mass and
mixing since neutrinos are produced and detected in the form of
weak-interaction eigenstates whereas their motion as they propagate
from the point of production to their detection is dictated by the
mass eigenstates \cite{pcrv}. In the two-generation formalism, the
mixing probability is:
\begin{equation}
P(\nu_1 \rightarrow \nu_2) = \sin^2 2\alpha \sin^2 \left(\frac{1.27
\Delta m^2 L}{E_\nu}\right)
\label{eq:posc}
\end{equation}
where $\Delta m^2$ is the mass squared difference of the mass
eigenstates in ${\rm eV^2}$, $\alpha$ is the mixing angle, $E_\nu$ is
the incoming neutrino energy in GeV, and $L$ is the distance between
the point of creation and detection in km.

To date the best limits from accelerator experiments for $\nu_\mu
\rightarrow \nu_e$ oscillations come from fine-grained calorimetric
(e.g.: BNL-E734 \cite{e734}, BNL-E776 \cite{e776}) or fully active
detectors (e.g. KARMEN \cite{karm}, LSND \cite{lsnd}) searching for
quasi-elastic charged current production of electrons. The LSND
experiment, using a liquid scintillator neutrino target, has reported
a signal consistent with $\bar{\nu}_\mu \rightarrow \bar{\nu}_e$
oscillations at a $\sin^2 2\alpha \approx 10^{-2}$ and $\Delta m^2
\stackrel{>}{\scriptstyle\sim} 1$~eV${^2}$ \cite{lsnd}. The CCFR
collaboration has previously reported a limit on $\nu_\mu \rightarrow
\nu_e$ oscillations using the ratio of neutral to charged current
neutrino events comparable in sensitivity to the above mentioned
limits \cite{donna}.

In this report we present new limits on $\nu_\mu \rightarrow \nu_e$
oscillations based on the statistical separation of $\nu_e N$ charged
current interactions.

The CCFR detector \cite{ws90,bk91} consists of an 18~m long, 690~ton
total absorption target calorimeter with a mean density of ${\rm 4.2
g/cm^3}$, followed by a 10~m long iron toroidal spectrometer. The
target consists of 168 steel plates, each ${\rm 3 m \times 3 m \times
5.15 cm}$, instrumented with liquid scintillation counters placed
every two steel plates and drift chambers spaced every four plates.
The separation between scintillation counters corresponds to 6
radiation lengths, and the ratio of electromagnetic to hadronic
response of the calorimeter is $1.05$. The toroid spectrometer is not
directly used in this analysis which is based on the shower profiles
in the target-calorimeter.

\begin{figure}
\centerline{
\psfig{figure=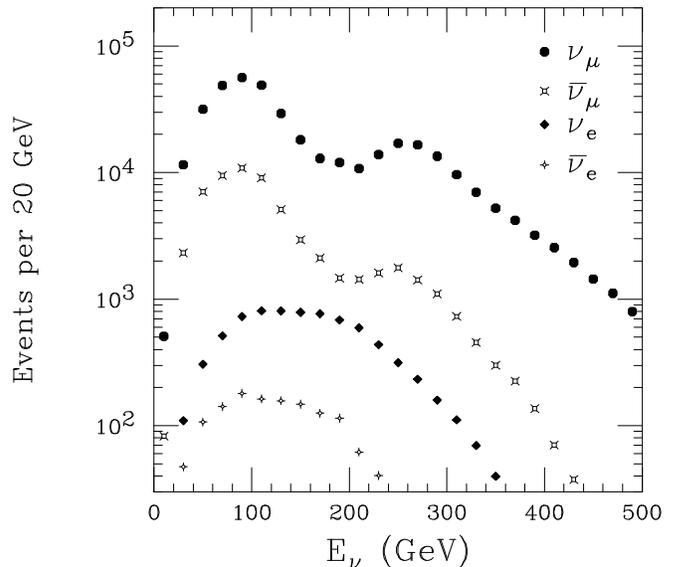,width=\columnwidth}}
\caption{Neutrino energy spectra for $\nu_\mu$, $\overline\nu_\mu$,
$\nu_e$, and $\overline\nu_e$ at the CCFR detector for the FNAL
wideband neutrino beam (Monte Carlo based on relative $\nu_\mu$ and
$\overline\nu_\mu$ fluxes).}
\label{fig:enu}
\end{figure}

The Fermilab Tevatron Quadrupole Triplet neutrino beam is a
high-intensity, non-sign-selected wideband beam with a $\nu$ :
$\overline\nu$ flux ratio of about 2.5 : 1 and usable neutrino energies
up to 600 GeV. The production target is located 1.4 km upstream of the
neutrino detector and is followed by a 0.5 km decay region. The resulting
neutrino energy spectra for $\nu_\mu$, $\overline\nu_\mu$, $\nu_e$,
and $\overline\nu_e$ induced events are shown in Figure \ref{fig:enu}.
The beam contains a 2.3\% fraction of electron neutrinos, 82\% of
which are produced from $K^\pm \rightarrow \pi^0 e^\pm
\stackrel{_{(-)}}{\nu_e}$.

The neutrino interactions observed in the detector can be divided into
three classes depending on the type of the incoming neutrino and on
the interaction type:
\begin{enumerate}
\item  $\nu_{\mu}N \rightarrow \mu^-X$ ($\nu_{\mu}$ charged current (CC)
events).
\item $\nu_{\mu,e}N \rightarrow \nu_{\mu,e}X$ ($\nu_{\mu,e}$ neutral
current (NC) events).
\item $\nu_e N \rightarrow eX$ ($\nu_e$ CC events).
\end{enumerate}

All three types of neutrino interactions initiate a cascade of hadrons
that is registered by the drift chambers and scintillation counters.
The $\nu_\mu$~CC events are characterized by the presence of a muon
produced in the final state which penetrates beyond the end of the
hadron shower, depositing energy characteristic of a minimum ionizing
particle \cite{ws90} in a large number of consecutive scintillation
counters. Conversely, the electron produced in a $\nu_e$~CC event
deposits energy in a few counters immediately downstream of the
interaction vertex which changes the energy deposition profile of the
shower. The electromagnetic shower is typically much shorter than the
hadron shower and the two cannot be separated for a $\nu_e$~CC event.

In this analysis four experimental quantities are calculated for each
event: the length, the transverse vertex position, the visible energy
and the shower energy deposition profile. The event length is
determined to be the number of scintillation counters spanned from the
event vertex to to the last counter with a minimum-ionizing pulse
height. The mean position of the hits in the drift chamber immediately
downstream of the interaction vertex determines the transverse vertex
position. The visible energy in the calorimeter, $E_{vis}$ is obtained
by summing the energy deposited in the scintillation counters from the
interaction vertex to five counters beyond the end of the shower. The
shower energy deposition profile is characterized by the ratio of the
sum of the energy deposited in the first three scintillation counters
to the total visible energy. Accordingly, we define
\begin{equation}
\eta_3 = 1 - \frac{E_1 + E_2 + E_3}{E_{vis}} \label{eq:eta3}
\end{equation}
where $E_i$ is the energy deposited in the $i^{th}$ scintillation
counter downstream of the interaction place.

The most downstream counter with energy deposited from the products of
the neutrino interaction (CEXIT) occurs at the end of the hadron
shower for $\nu_\mu$~NC and $\nu_e$~CC events but is determined by the
muon track for most $\nu_\mu$~CC events. We isolate the events without
a muon track by requiring CEXIT to be no more than 10 counters
downstream from the end of the hadron shower. We parametrize the event
length which contains 99\% of such events as:
\begin{equation}
L_{NC} = 4.+ 3.81\times \log(E_{vis})
\end{equation}

In order to measure the number of $\nu_e$~CC events we divide the
neutrino events into two classes: ``short'' if they deposit energy
over an interval shorter than $L_{NC}$, and ``long'' otherwise. The
long events consist almost exclusively of class~1 events, while the
short ones are a mixture of class~2, class~3 and class~1 events with a
low energy muon which cannot be separated on an event-by-event basis.

Based on Lund studies, we take the hadron showers produced in NC and
CC interactions to be the same. Any difference in the shower energy
deposition profile of long and short events is attributed to the
presence of $\nu_e$~CC interactions in the short sample. To compare
directly the long and short events a muon track from the data was
added to the short events to compensate for the absence of a muon in
NC events. The fraction, {\em f}, of $\nu_\mu$~CC events with a low
energy muon contained in the short sample which now have two muon
tracks was estimated from a detailed Monte Carlo of the experiment in
the range of 20\%. A simulated sample of such events was obtained by
choosing long events with the appropriate energy distribution from the
data to which a second short muon track was added in software. The
length of the short track and the angular distribution were obtained
from a Monte Carlo of $\nu_\mu$~CC events.

To simulate $\nu_e$ interactions in our detector we assume $\nu_\mu -
\nu_e$ universality. The electron neutrino showers were generated by
adding a GEANT \cite{g321} generated electromagnetic shower of the
appropriate energy to events in the long data sample. The energy
distribution of the electron neutrinos and the fractional energy
transfer $y$ were generated using a detailed Monte Carlo simulation of
the experiment. Since the hadron showers in the long sample already
have a muon track, the $\nu_e$ sample can be compared directly with
the short and long events.

\begin{figure}
\centerline{
\psfig{figure=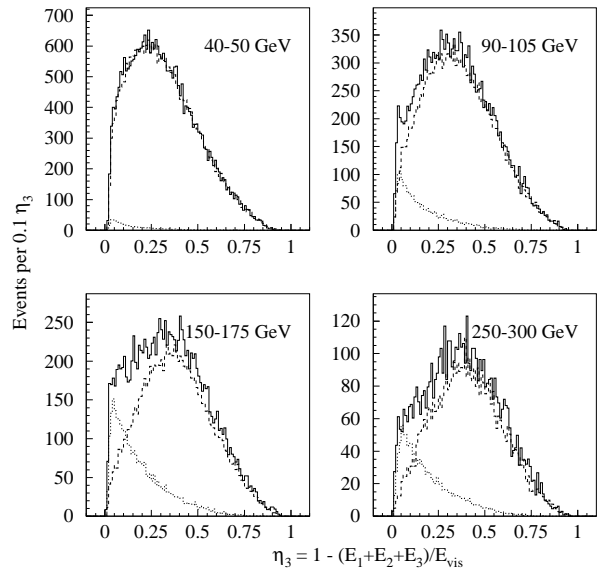,width=\columnwidth}}
\caption{Eta distributions for short (solid line), long (dashed line)
and $\nu_e$ (dotted line) events in four of the energy bins studied.
The $\nu_e$ and long distributions are normalized to the respective
number of events predicted by the fit.}
\label{fig:etas}
\end{figure}

The long and short $\eta_3$ distributions were further corrected by
subtracting the contamination due to cosmic ray events. The cosmic ray
background was estimated from the event sample collected during a beam
off gate using an identical analysis procedure as for the data gates.
Additionally, the $\eta_3$ distribution of short $\nu_\mu$~CC events,
normalized to the predicted fraction {\em f}, was subtracted from the
short event sample. The $\eta_3$ distributions for short, long, and
$\nu_e$ events for various energy bins are shown in
Figure~\ref{fig:etas}.

For this oscillation search we measure the absolute flux of $\nu_e$'s
at the detector and compare it to the flux predicted by a detailed
beamline simulation \cite{ca94}. Any excess could be interpreted as a
signal of $\nu_\mu \rightarrow \nu_e$ oscillations. The $\nu_\mu$ flux
was determined directly from the low hadron energy CC event sample,
normalized to the total neutrino cross-section \cite{flux}. The same
beamline simulation is used to tag the creation point of each
simulated $\nu_\mu$ along the decay pipe, and give the number of
predicted $\nu_\mu$'s at the detector normalized to the number
observed at the detector divided by $1-P(\nu_\mu \rightarrow \nu_e)$.
$P(\nu_\mu \rightarrow \nu_e)= P(\overline\nu_\mu \rightarrow
\overline\nu_e)$ is the oscillation probability determined from
eq.~(\ref{eq:posc}), assuming CP invariance. The predicted electron
neutrino flux is normalized to the {\em produced} number of
$\nu_\mu$'s. The $\nu_e$ flux from neutrino oscillations is calculated
by multiplying the {\em produced} number of $\nu_\mu$'s by $P(\nu_\mu
\rightarrow \nu_e)$.

The events selected are required to deposit a minimum of 30 GeV in the
target calorimeter to ensure complete efficiency of the energy
deposition trigger. Additionally, we require the event vertex to be
more than 5 counters from the upstream end of the target and five
counters plus the separation length from the downstream end and less
than $50"$ from the detector centre-line. The resulting data sample
consists of 632338 long events and 291354 short ones.

To extract the number of $\nu_e$~CC events in each of 15 $E_{vis}$
bins, we fit the corrected shape of the observed $\eta_3$ distribution
for the short sample to a combination of $\nu_\mu$~CC and $\nu_e$~CC
distributions with appropriate muon additions:
\begin{equation}
{\rm \nu_\mu NC (+ \mu) = \alpha \: \nu_\mu CC + \beta \: \nu_e CC (+
\mu)}
\end{equation}

\begin{figure}
\centerline{
\psfig{figure=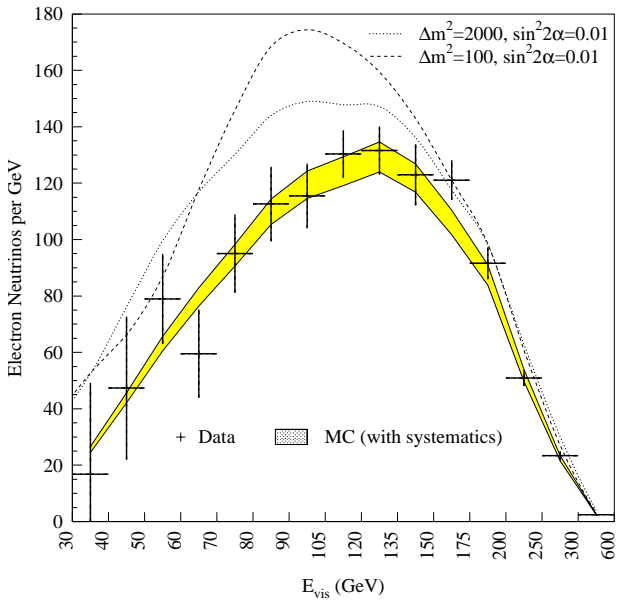,width=\columnwidth}}
\caption{Number of electron neutrinos as a function of visible energy.
For electron neutrinos the visible energy is equal to the total
neutrino energy. The filled band shows Monte Carlo prediction assuming
no oscillations. The dotted curve corresponds to $\nu_\mu \rightarrow
\nu_e$ oscillations with $\Delta m^2 = 2000$~${\rm eV^2}$ and $\sin^2
2 \alpha = 0.01$ and the dashed curve to $\Delta m^2 = 100$~${\rm
eV^2}$ and $\sin^2 2 \alpha = 0.01$}
\label{fig:result}
\end{figure}

The $\chi^2$ of the fit in each of the 15 $E_{vis}$ bins ranges from
33.2 to 77.7 for 41 degrees of freedom (DoF) with a mean value of
48.4. Figure \ref{fig:result} shows that the measured number of
$\nu_e$~CC's agrees with the Monte Carlo prediction in each energy
bin. The $\chi^2$ value with a no-oscillations assumption is $9.97/15$
DoF.

The major sources of uncertainties in the comparison of the electron
flux extracted from the data to that predicted by the Monte Carlo are:
(i) The statistical error from the fit in the extraction of the
$\nu_e$ flux. (ii) The error in the shower shape modeling, estimated
by extracting the $\nu_e$ flux using two definitions of $\eta$.
Analogous to the definition of $\eta_3$ given in eq.~(\ref{eq:eta3}),
we define $\eta_4$ to be the ratio of the sum of the energy deposited
outside the first four scintillation counters to the total visible
energy. If the modeling of the showers were correct, the difference in
the number of electron neutrinos measured by the two methods should be
small, any difference is used to estimate the systematic error. Since
this error was shown not to be correlated among energy bins, we add it
in quadrature to the statistical error from the fit and take this to
be the combined basic error. (iii) The 1\% uncertainty in the absolute
energy calibration of the detector changes the relative neutrino flux
which is extracted using the subset of the data sample with low hadron
energy \cite{flux}. (iv) The uncertainty in the incident flux of
$\nu_e$'s on the detector is estimated to be $4.1\%$ \cite{ca94}. This
error is dominated by a 20\% production uncertainty in the $K_L$
content of the secondary beam which produces 16\% of the $\nu_e$ flux.
The majority of the $\nu_e$ flux comes from $K_{e_{3}}^\pm$ decays,
which are well-constrained by the observed $\nu_\mu$ spectrum from
$K_{\mu_{2}}^\pm$ decays \cite{ca94}. Other sources of systematic
errors were also investigated and found to be small.

\begin{table}
\caption{The result for $\sin^2 2\alpha$ from the fit at each $\Delta
m^2$ for $\nu_\mu \rightarrow \nu_e$ oscillations. The 90\% C.L.
upper limit is equal to the best fit $\sin^2 2\alpha +
1.28\sigma$.}
\label{tab:bestfit}
\begin{center}
\begin{tabular}{cccccc}
$\Delta m^2$ (eV$^2$) & Best fit & $\sigma$ & $\Delta m^2$ (eV$^2$) &
Best fit & $\sigma$ \\
\tableline
\begin{tabular}{r}
   1.0  \\
    2.0  \\
    3.0  \\
    4.0  \\
    5.0  \\
    7.0  \\
    9.0  \\
   10.0  \\
   20.0  \\
   30.0  \\
   40.0  \\
   50.0  \\
   60.0  \\
   70.0  \\
   80.0  \\
   90.0  \\
  100.0  \\
  125.0  \\
  150.0  \\
\end{tabular} & \begin{tabular}{r}
-0.1741 \\
-0.0501 \\
-0.0153 \\
-0.0112 \\
-0.0051 \\
-0.0036 \\
-0.0021 \\
-0.0023 \\
-0.0004 \\
-0.0003 \\
-0.0002 \\
-0.0002 \\
-0.0002 \\
-0.0002 \\
-0.0003 \\
-0.0003 \\
-0.0002 \\
 0.0004 \\
 0.0005 \\
\end{tabular} & \begin{tabular}{r}
1.6501 \\
0.4107 \\
0.1852 \\
0.1041 \\
0.0671 \\
0.0345 \\
0.0213 \\
0.0173 \\
0.0048 \\
0.0026 \\
0.0018 \\
0.0015 \\
0.0014 \\
0.0014 \\
0.0014 \\
0.0015 \\
0.0015 \\
0.0018 \\
0.0019 \\
\end{tabular} & \begin{tabular}{r}
  175.0  \\
  200.0  \\
  225.0  \\
  250.0  \\
  275.0  \\
  300.0  \\
  350.0  \\
  400.0  \\
  450.0  \\
  500.0  \\
  600.0  \\
  700.0  \\
  800.0  \\
 1000.0  \\
 1500.0  \\
 2000.0  \\
 5000.0  \\
10000.0  \\
20000.0  \\
\end{tabular} & \begin{tabular}{r}
 0.0000 \\
-0.0002 \\
-0.0003 \\
-0.0004 \\
-0.0004 \\
-0.0004 \\
-0.0004 \\
-0.0003 \\
-0.0003 \\
-0.0004 \\
-0.0005 \\
-0.0003 \\
-0.0002 \\
-0.0004 \\
-0.0003 \\
-0.0004 \\
-0.0003 \\
-0.0004 \\
-0.0004 \\
\end{tabular} & \begin{tabular}{r}
0.0016 \\
0.0014 \\
0.0013 \\
0.0012 \\
0.0012 \\
0.0012 \\
0.0012 \\
0.0013 \\
0.0015 \\
0.0016 \\
0.0019 \\
0.0018 \\
0.0018 \\
0.0017 \\
0.0017 \\
0.0017 \\
0.0018 \\
0.0017 \\
0.0017 \\
\end{tabular}
\end{tabular}
\end{center}
\end{table}

The data are fit by forming a $\chi^2$ which incorporates the Monte
Carlo generated effect of oscillations, the basic error, and terms
with coefficients accounting for systematic uncertainties. A best fit
$\sin^2 2\alpha$ is determined for each $\Delta m^2$ by minimizing
the $\chi^2$ as a function of $\sin^2 2\alpha$ and these systematic
coefficients. At all $\Delta m^2$, the data are consistent with no
observed $\nu_\mu \rightarrow \nu_e$ oscillations. The statistical
significance of the best-fit oscillation at any $\Delta m^2$ is at
most $0.3 \sigma$.

\begin{figure}
\centerline{
\psfig{figure=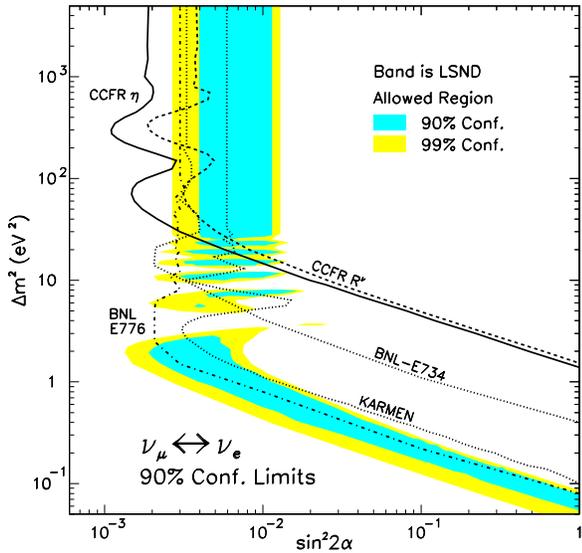,width=\columnwidth}}
\caption{Excluded region of $\sin^2 2\alpha$ and $\Delta m^2$ for
$\nu_\mu \rightarrow \nu_e$ oscillations from this analysis at 90\%
confidence is the area to the right of the dark, solid curve.}
\label{fig:osc}
\end{figure}

The frequentist approach \cite{pdg} is used to set a 90\% confidence
upper limit for each $\Delta m^2$. The limit in $\sin^2 2\alpha$
corresponds to a shift of 1.64 units in $\chi^2$ from the minimum
$\chi^2$ (at the best fit value in Table~\ref{tab:bestfit}). 
The 90\% confidence upper limit is plotted in Figure \ref{fig:osc} for
$\nu_\mu \rightarrow \nu_e$. The best limit of $\sin^2 2\alpha < 1.1
\times 10^{-3}$ is at $\Delta m^2 = 300$~${\rm eV^2}$. For $\sin^2
2\alpha = 1$, $\Delta m^2 > 1.6$~${\rm eV^2}$ is excluded, and for
$\Delta m^2 \gg 1000$~${\rm eV^2}$, $\sin^2 2\alpha > 1.8 \times
10^{-3}$.

Under the assumption that there are no oscillations, this data can
also be used to test $\nu_\mu (\bar\nu_\mu) \leftrightarrow \nu_e
(\bar\nu_e)$ universality by comparing the observed $\nu_e$ flux to
that predicted by the Monte Carlo. From this comparison we determine
the ratio of the cross sections averaged over our flux to be
$\sigma_{CC}(\nu_\mu)/\sigma_{CC}(\nu_e) = 1.026 \pm 0.055$. This is
currently the most stringent test of universality at high space-like
momentum transfer.

In conclusion, we have used the difference in the longitudinal shower
energy deposition pattern of $\nu_e N$ versus $\nu_\mu N$ interactions
to search for $\nu_\mu \rightarrow \nu_e$ oscillations with a
coarse-grained calorimetric detector. We see a result consistent with
no neutrino oscillations and find 90\% confidence level excluded
regions in the $\sin^2 2\alpha - \Delta m^2$ phase space. This result
is the most stringent limit to date for $\nu_\mu \rightarrow \nu_e$
oscillation for $\Delta m^2 > 25$~${\rm eV^2}$. We also tested
$\nu_\mu (\bar\nu_\mu) \leftrightarrow \nu_e (\bar\nu_e)$ universality
and found the ratio of the $\nu_\mu$-to-$\nu_e$ cross-section to be
$1.026 \pm 0.055$.

\end{document}